\documentclass{ws-procs975x65}

\begin{document}

\title{Interacting quintessence and the coincidence problem}

\author{W. Zimdahl}

\address{Institut f\"ur Theoretische Physik, \\
Universit\"at zu  K\"oln,
D-50937 K\"oln, Germany \\
E-mail: zimdahl@thp.uni-koeln.de}

\author{D. Pav\'{o}n}
\address{Departamento de F\'{\i}sica, \\
Universidad Aut\'{o}noma de Barcelona, 08193 Bellaterra
(Barcelona), Spain}

\author{L.~P. Chimento and A.~S. Jakubi}

\address{Departamento de F\'{\i}sica, \\
Universidad de Buenos Aires, 1428 Buenos Aires, Argentina}

\maketitle \abstracts{ We investigate the role of a possible
coupling of dark matter and dark energy. In particular, we explore
the consequences of such an interaction for the coincidence
problem, i.e., for the question, why the energy densities of dark
matter and dark energy are of the same order just at the present
epoch. We demonstrate, that, with the help of a suitable coupling,
it is possible to reproduce any scaling solution $\rho_X \propto
\rho_M a^\xi$, where $a$ is the scale factor of the
Robertson-Walker metric and $\xi$ is a constant parameter.
$\rho_X$ and $\rho_M$ are the densities of dark energy and dark
matter, respectively. Furthermore, we show that an interaction
between dark matter and dark energy can drive the transition from
an early matter dominated era to a phase of accelerated expansion
with a stable, stationary ratio of the energy densities of both
components.}

\section{Introduction}

Our Universe seems to be made of roughly two third of something
which is now called dark energy (DE) and of about one third of
mainly cold dark matter (CDM). Usually, these components are
considered as independent ingredients of the cosmic substratum.
Given, that we neither know the nature of DE nor that of CDM, one
cannot exclude that there exists a coupling between them. Such a
situation is even the more general one compared with the uncoupled
case. Here we explore some of the consequences that different
types of interactions may have for the dynamics of the Universe.

\section{Scaling solutions}

A phenomenological generalization of the $\Lambda$CDM model,
characterized by $\rho_M \propto a^{-3}$ and  $\rho_X
\equiv\rho_\Lambda = {\rm const} $, is\cite{dalal}
\begin{equation}
\frac{\rho _{M}}{\rho _{X}} = r_0\left(\frac{a _{0}}{a}
\right)^{\xi } \ ,\label{1}
\end{equation}
with a dark energy equation of state parameter $w_X \equiv p_X
/\rho_X$ which is assumed to be a negative constant. For $\xi = 3$
one obviously recovers the $\Lambda$CDM model. The limit $\xi = 0
$ corresponds to stationary ratio of the energy densities. For any
value $\xi < 3 $ the coincidence problem is less severe than for
the $\Lambda$CDM model. Here we want to show, that $\xi < 3 $ can
be achieved by a suitable interaction according to
\begin{equation}
\dot{\rho}_{M} + 3H \rho_{M} = Q \ , \quad \dot{\rho}_{X} + 3H
\left(1+w _{X}\right)\rho _{X} = -Q \ .
\label{2}
\end{equation}
Scaling solutions of the type (\ref{1}) are then obtained for a
choice
\begin{equation}
Q  = -3H\frac{\frac{\xi }{3} + w _{X}} {1 + r_0 \left(1+z
\right)^{\xi }}\,\, \rho _{M}\ .
\end{equation}
This demonstrates that by a suitable interaction one may reproduce
any desired scaling behavior of the energy densities.\cite{GRG}
The  uncoupled case is recovered for $ \xi + 3w _{X} = 0$. For any
$\xi + 3w _{X} \neq 0$ we obtain an alternative cosmological
model.

\subsection{The case $w_X = -1$, $\xi = 1$}

In such a case we have $Q>0$, i.e., a continuous energy transfer
from the $X$-component to the matter fluid and a total energy
density
\begin{equation}
\rho  = \frac{\rho _{0}}{\left(1+r_0 \right)^3} \left[1 +
r_0\frac{a_0}{a}\right]^{3}\ .
\end{equation}
Compared with the corresponding expression of the $\Lambda$CDM
model, $\rho_{(\Lambda CDM) } \propto \rho _{0} \left[1 +
r_0\left(a_0 /a\right)^{3}\right]$, the sum of the powers
($\Lambda$CDM) is replaced by the power of a sum. For the
luminosity distance of this model we obtain an expression which
differs from the corresponding one of the $\Lambda$CDM model only
in the third order in the redshift. While the results of both
models are indistinguishable from the current SN Ia data, the
coincidence problem is less severe in the interacting
model.\cite{GRG}

\subsection{The stationary case $\xi = 0$}

This case has $\rho _{X}$, $\rho _{M} \propto a ^{-\nu }$, where
$\nu = 3 \left(1 + r_0 + w_X\right)/\left(1 + r_0\right)$ and $a
\propto t^{2/\nu}$. There is accelerated expansion for $ \nu < 2$,
corresponding to $3w_X < - \left (1+r_0\right)$. Modelling the $X$
component by a scalar field, the corresponding potential is
\begin{equation}
V(\phi) \propto\exp{\left[- \kappa \phi \right]}\ ,\qquad \kappa =
\sqrt{\frac{24\pi G}{\left(1+w_X\right) \left(1+r_0\right)}}
\left[1 + w_X + r_0 \right]\ .
\end{equation}
The condition for accelerated expansion translates into $\kappa
^{2} < 24\pi G \frac{w_X^{2}}{\left(1+r_0 \right) \left(1+w_X
\right)}$. We conclude that there exists a specific interaction
which admits a stationary ratio of the energy densities and is
compatible with an accelerated expansion of the
Universe.\cite{PLB} In a next step we outline, how such a state
might be dynamically approached.

\section{Asymptotically stable solution }

With an ansatz $Q = 3H c^{2}\rho$ for the interaction, equations
(\ref{2}) admit two stationary solutions, $r_s^\pm$, for the ratio
$r\equiv \rho_M /\rho_X$ of the energy densities. These have the
properties $r_s^+ - r_s^- \geq 0$ and  $r_s^+ r_s^- = 1$. For $r =
r_s^+ $ we have $\rho_M
> \rho_X$, i.e., matter dominance, while for $r = r_s^- $ the
reverse relation $\rho_M < \rho_X$ holds, equivalent to a dark
energy dominated state. The dynamics of $r$ in terms of these
stationary solutions is given by
\begin{equation}
\dot r=-3c^2H\left[(r-r_s^-)(r_s^+ -r)\right]\ .
\end{equation}
While $r_s^+ $ can be shown to be unstable, $r_s^-$ is a stable,
stationary state. There exists a solution
\begin{equation}
r(y)=r_s^-\frac{1 + yr_s^+} {1+yr_s^-} \ ,\qquad r\left(1\right) =
1 \ ,
\end{equation}
with $y= \left(a_{eq}/a\right)^{\lambda}$ and $\lambda = -3w_X
\left(1-r_s^-\right)/\left(1 + r_s^-\right)$, according to which
the density ratio evolves from the unstable stationary value
$r_s^+$ for $y \gg 1$ (matter dominance) to the stable, stationary
state $r_s^-$ for $y\ll 1 $ (dark energy dominance).\cite{PRD} The
deceleration parameter decreases with the expansion. In
particular, it may switch from positive to negative values.
Assuming as an example $w_X = -1$, $r_0  \approx 3/7$ and, say,
$r_s^- \approx 1/7 $, we find that the value $r_{acc}$ of $r$ at
which this transition occurs is $r_{acc}\approx 2$. This
corresponds to a redshift of $z_{acc} \approx 1.6$.

\section*{Acknowledgments}
This work was supported by the Deutsche Forschungsgemeinschaft, by
grant BFM 2003-06033 of the Spanish Ministry of Science and
Technology and by the University of Buenos Aires under Project
X223.


\end{document}